\documentclass[twocolumn,aps,floats,superscriptaddress]{revtex4}
\usepackage{graphicx,epsfig}
\usepackage{times}
\usepackage{graphics,dcolumn,bm,float}
\usepackage{amssymb,amsmath,rotate,color}
\usepackage[normalem]{ulem}
\usepackage{soul}

\begin{document}

\title{Mechanisms of jump to contact and conductance plateau formation in copper atomic junctions in vacuum and aqueous environments}
\author{Alireza Saffarzadeh}
\altaffiliation{Author to whom correspondence should be addressed.
Electronic mail: asaffarz@sfu.ca} \affiliation{Department of Physics, Payame Noor University, P.O. Box 19395-3697 Tehran, Iran}
\affiliation{Department of Physics, Simon Fraser University,
Burnaby, British Columbia, Canada V5A 1S6}
\author{George Kirczenow}
\affiliation{Department of Physics, Simon Fraser University,
Burnaby, British Columbia, Canada V5A 1S6}

\date{\today}

\begin{abstract}
The interplay between groups of water molecules and single-atom contacts, as reflected in the electrical conductances and mechanical forces of copper atomic junctions, is explored by means of first-principles theory and semi-empirical calculations. We study the influence of the atomic geometries of copper electrodes with pyramidal and non-crystalline structures in the presence and absence of water on the conductance profiles as the electrodes approach each other. It is shown that the atomic arrangements of nano-contacts have crucial effects on the formation of plateaus and the conductance values. Groups of hydrogen bonded water molecules bridge the junction electrodes before a direct Cu-Cu contact between the electrodes is made. However, the bridging of the two copper electrodes by a single H$_{2}$O molecule only occurs in the junctions with pyramidal electrodes. Our findings reveal that the presence of H$_{2}$O molecules modifies strongly the conductance profile of these junctions. 
In the absence of water molecules, the pyramidal junctions exhibit continuous transitions between integer conductance plateaus, while in the presence of H$_{2}$O molecules, these junctions show abrupt jump to contact behavior and no well-defined conductance plateaus.
 By contrast, in the absence of H$_{2}$O molecules, the non-crystalline junctions display jump to contact behavior and no well-defined plateaus, while in the presence of H$_{2}$O molecules they exhibit a jump to contact and abrupt transitions between fractional and integer plateaus. 
 
\end{abstract}
\maketitle

\section{Introduction}
Metallic junctions and the process of their contact formation have been the subject of many experimental and theoretical studies exploring physical properties, such as  electrical conduction \cite{Agrait,Vardimon2014,Fernandez2016,Calvo2018}, quantum interference \cite{Smit2003,Vega,Aiba1} and thermoelectric energy conversion \cite{Tsutsui1,Mosso,Aiba1,AFG2018} at the atomic and molecular scales. Achieving control over the electronic transport properties of single atom contacts is crucial for the realization of practical nanoelectronic devices. In the context of single atomic junctions, when the distance between two metal electrodes is sufficiently large, a tunnel current that increases with decreasing electrode separation is observed. As the electrodes are brought closer together, at a specific separation a jump to contact usually occurs. This is followed by a plateau in the conductance \cite{Gimzewski1987} that can be observed by means of conductance measurement techniques, including scanning tunneling microscopy (STM) and mechanically controllable break junction (MCBJ) methods. However, the jump does not always occur and, instead, the current can increase smoothly from tunneling to the metallic contact regime \cite{Cross1998,Halbritter2003,Limot2005,Untiedt2007}. In fact, this phenomenon depends on the type of metal and the geometry of the two electrodes \cite{Untiedt2007,Kroger2008,Kroger2009}.

The presence of appropriate impurities, adsorbates and gaseous environments in the growth process can strongly affect the formation and stability of metallic junctions. Adsorbed gas atoms or molecules may cause a deviation from integer values in the conductance at room temperature even at ultrahigh vacuum conditions.  For this reason, the influence of light gas atoms and molecules on the structural and electronic properties of metallic junctions and point contacts has been widely studied experimentally \cite{Thijssen2008,Thijssen2006,Li2016,Smit2002,Djukic2006,Christlieb2002,Djukic2005,Csonka2004,Kiguchi2007,Nakazumi2010,Li2015} and theoretically \cite{Thygesen2005,Novaes2006,Cakir2011,Amorim2012,Zheng2015,Duan2017}. For instance, using the MCBJ technique it has been shown that oxygen atoms can be incorporated in noble metal (Au, Ag, Cu) atomic chains, leading to a considerable reduction in conductance for Ag and Cu chains and retaining the conductance near one quantum unit ($1g_{0}=2e^{2}/h$) for Au atomic chains\cite{Thijssen2008,Thijssen2006}. The conductance of a single hydrogen molecule bridging two metal electrodes such as Pt, Pd, Co, Au, Ag, and Cu has also been studied  \cite{Djukic2006,Christlieb2002,Djukic2005,Thygesen2005,Smit2002,Csonka2004,Kiguchi2007,Nakazumi2010,Li2015}. In some experiments, the histogram of conductance values has shown additional structures with conductances lower than $1g_{0}$, suggesting that various atomic configurations of the hydrogen molecule between electrodes may form, depending on the geometrical symmetry of the electrodes \cite{Kiguchi2007,Li2015}.  Thus, measured conductance values near 0.3$g_{0}$ for Cu electrodes after the introduction of hydrogen have been reported, while variable conductance values below $1g_{0}$ for Au and Ag electrodes exposed to hydrogen have been observed \cite{Li2015}.

Despite the aforementioned experimental and theoretical studies of the interaction of O$_{2}$ and H$_{2}$ molecules with metallic atomic junctions, only a few theoretical and experimental investigations of metallic junctions and nanocontacts in the presence of H$_{2}$O molecules have been reported so far \cite{Tal2008,Barzilai2013,Li-PCCP2015,Fukuzumi2018}. Although the water molecule is relatively small, it plays an important role in many processes, such as corrosion, catalysis, electrolysis, photosynthesis, and in hydrogen fuel cells. Most of the investigations considered only the cases in which the atomic junction was stretched until it ruptured. An alternative situation is to bring the electrodes together (as in STM experiments where the tip is brought into contact with the sample) and to measure the conductance as the distance between the electrode tips of the junction is reduced. 

In order to shed light on the latter situation we consider copper atomic junctions that have attracted attention previously \cite{Xu2015,Nam2016} due to their possible applications in nano-electrical systems. For instance, copper nanowires have ultra-low junction resistance which can be utilized for interconnects in future nanoscale devices \cite{Xu2015} and for high-performance and low-haze transparent conductors \cite{SRYe2014,FCui2015}.
Here the conductance of these junctions in the presence and absence of groups of H$_{2}$O molecules is explored theoretically by means of \textit{ab initio} calculations and a semi-empirical tight-binding method based on the extended H\"{u}ckel theory \cite{Ammeter,YAEHMOP,George2010}. We have selected H$_{2}$O molecules as a common contaminant which can exist in the experimental equipment. 

In experiments the metal junction is usually stretched until it breaks and then the metal tips that are formed are brought back together and the junction is re-formed. This is repeated many times in typical experiments. The stretching and breaking of copper junctions in the presence of water molecules was studied in Ref. \onlinecite{Li-PCCP2015}. Here we consider the process where the junction is being re-formed after being broken. At the last stage of stretching of the junction before it breaks a bridge consisting of a single metal atom or a chain of metal atoms usually forms between the two metal electrodes of the junction. When such an atomic bridge breaks it is reasonable to expect one or both of the metal tips that remain to terminate in a single metal atom, at least in some cases. Here we study tips that terminate in this way. However, it is not known at this time what degree of crystalline order may be present in the two electrodes in the immediate vicinity of the tip atoms after the junction has been reformed and broken many times. Thus we consider two cases that may be representative: (i) Pyramidal electrodes that are perfectly ordered and (ii) non-crystalline electrodes whose geometries are strongly disordered although they correspond to a local energy minimum. Examples of individual metal electrodes of these two types are shown in Fig. 1 (a) and (b), respectively.

We show that the atomic arrangement and structural symmetry of the copper electrodes are crucial in the formation and stability of conductance plateaus as the junction length is reduced. The conductance profiles exhibit significant changes due to the presence of water molecules that hinder the tip Cu atoms coming into contact. The introduction of water molecules can result in a change from a smooth transition from tunneling to the metallic contact regime in the absence of water, to a sudden jump to contact in the presence of water, depending on the atomic arrangement in the electrodes. Regardless of the geometry of the electrodes, the conductance value is always below $1g_{0}$ when the Cu tip atoms come into a contact in the presence of water molecules. Moreover, we show that force measurements can detect the jump to contact in noncrystalline electrodes and reveal the onset of bridging of the two pyramidal electrodes by a single water molecule. The forces responsible for the jump to contact are found to be short ranged and are due to interatomic bond formation and bond rearrangement. A variety of detailed mechanisms responsible for the jump to contact are identified.

We have organized the paper as follows. In Sec. II, the computational details of the energy optimization of atomic
junctions and of the electronic transport calculations are presented. In Sec. III, we discuss the conductance and mechanical force results for copper junctions with both pyramidal and non-crystalline geometries for the electrodes in the presence and absence of water molecules.
Finally, in Sec. IV, we conclude this work with a general discussion of the results and of the different  jump to contact mechanisms that we have identified.
\begin{figure}[ht]
\begin{center}
\includegraphics[scale = 0.48]{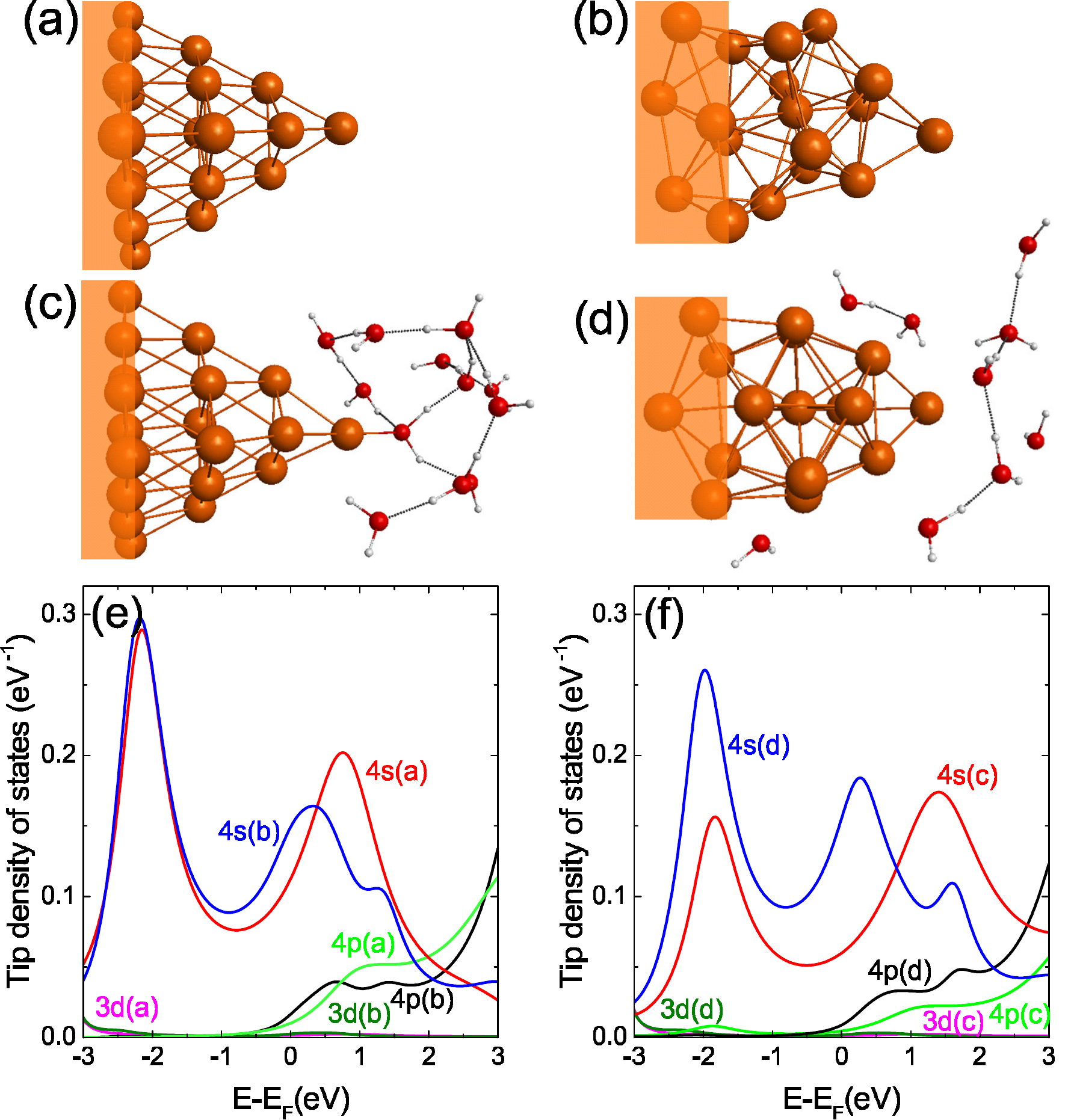}
\caption{(Color online) Optimized geometry of single electrodes with [(a), (c)] pyramidal and [(b), (d)] non-crystalline atomic configurations attached to semi-infinite leads (shaded area) in the [(a), (b)] absence and [(c), (d)] presence of water molecules. [(e), (f)] Calculated local density of states of 4$s$, 4$p$, and 3$d$ orbitals versus energy at tip atom of the single electrodes (a)-(d) in the (e) absence and (f) presence of water molecules. The orange, red, and grey atoms represent copper, oxygen, and hydrogen atoms, respectively. }\label{F1}
\end{center}
\end{figure}

\section{Theory}
In order to investigate electron conduction through copper atomic junctions in the presence of water molecules, we consider two different atomic arrangements for the Cu electrodes: (i) pyramidal and (ii) non-crystalline structures. To optimize the geometry of the electrodes, we minimize the total energy of the structures by means of density functional theory as implemented in the GAUSSIAN 16 package with the Perdew-Burke-Ernzerhof (PBE) functionals and Lanl2DZ effective core potential and basis sets \cite{gaussian,Pedrew}. We do not consider van der Waals interactions because our copper electrodes take the form of compact atomic clusters for which the cohesive forces between neighboring atoms that are in close contact with each other are much stronger than the van der Waals forces between well separated atoms. Also in the relaxed spatial arrangements hydrogen bonds hold the water molecules in place and the forces between the water molecules are dominated by dipole-dipole interactions which are much stronger than the van der Waals forces since the latter are due to quantum fluctuations.
The electronic energy and ionic forces of all optimized geometries were converged within $10^{-5}$ eV and 0.0008 eV/\AA , respectively.
In the pyramidal arrangement, each electrode consists of four atomic layers comprised of 1 atom (the tip atom), 3 atoms,  6 atoms, and 10 atoms (the outermost layer), respectively. Initially, the electrodes are assumed to have the geometrical configuration of the ideal fcc lattice in the crystal direction $<111>$ with nearest neighbor distance 2.50 {\AA}. All atoms in the outermost layer of each pyramidal electrode are frozen during the optimization process while the other atoms are free to move. 
In the case of electrodes with non-crystalline atomic structure, however, the electrode atoms are not arranged in a well-ordered structure initially and only a single outermost atom of each 19-atom electrode is fixed during the structural relaxation. Well separated non-crystalline electrodes are constructed by adding copper atoms to them one by one at random locations and allowing the atomic geometries to become fully optimized after each atom is added.   In both the pyramidal and non-crystalline cases, if water is present, the water molecules are initially distributed randomly in the space between the two well separated electrodes and in the vicinity of the tip atoms, prior to optimization. Energy optimization for the copper electrodes in the presence or absence of water molecules is performed for all junction lengths included in the conductance profile.  For this case, our starting point is a relaxed geometry with well-separated electrodes. The other structures that we consider are generated from this one by approaching the electrodes slightly towards each other and then freezing the positions of the outermost atoms as discussed above and relaxing the positions of the other atoms of the system. This procedure is repeated multiple times until the electrodes merge into each other gradually.

To calculate the electrical conductances of the atomic junctions with pyramidal (non-crystalline) structures, we couple the junctions to electron reservoirs. To do this, each one of the valence orbitals 4$s$, 4$p_{x}$, 4$p_{y}$, 4$p_{z}$, 3$d_{xy}$, 3$d_{xz}$, 3$d_{yz}$, 3$d_{z^2}$, and 3$d_{x^2-y^2}$ belonging to the ten (seven) farthest atoms of each electrode from the junction is connected to a one-dimensional ideal lead representing electron reservoirs (see Figs. 1(a)-(d)) \cite{AFG2018,Kirczenow2005,Dalgleish2006,Piva2008,Cardamone2010,AFG2013}. The electron transmission amplitudes $t_{ji}(E)$ through the system consisting of the electrodes and water molecules are obtained by solving the Lippmann-Schwinger equation 
\begin{equation}
|\psi^{\alpha}\rangle=|\phi^{\alpha}\rangle+G_{0}(E)W|\psi^{\alpha}\rangle\ .
\end{equation}
Here $|\phi^{\alpha}\rangle$ is an electron eigenstate of the $\alpha$th ideal semi-infinite one-dimensional lead that is decoupled from the copper atoms representing the electrodes, $G_{0}(E)$ is the Green's function of the decoupled system of the leads and the atomic junction, $W$ is the coupling between the electrodes and the ideal leads, and $|\psi^{\alpha}\rangle$ is the scattering eigenstate of the complete coupled system associated with the incident electron state $|\phi^{\alpha}\rangle$. We assume that the electronic state $i$ of a carrier with velocity $v_{i}$ is coming from the left lead while the electronic state $j$ of a carrier with velocity $v_{j}$ is transmitted to the right lead.
Due to the limitations of density functional theory for transport calculations (see Sections 4.7.1 and 4.7.2 of Ref. \onlinecite{George2010} and references therein), the semi-empirical extended H\"{u}ckel model  with the parameters of Ammeter \textit{et al.} \cite{Ammeter,YAEHMOP} are used to evaluate the Hamiltonian matrix elements and atomic valence orbital overlaps that enter the Green's function $G_{0}(E)$ in Eq. (1). It should be mentioned that this methodology involves no fitting to any experimental data relating to transport.

Using the electron transmission amplitudes, the zero bias conductance $g$ at the Fermi energy $\epsilon_{F}$ of the macroscopic electrodes used in conductance measurements is calculated at zero temperature (for various junction lengths) from the Landauer formula \cite{George2010}
\begin{equation}\label{g}
g(\epsilon_{F})=g_{0}\sum_{i,j}|t_{ji}(\epsilon_{F})|^2 \frac{v_j}{v_i}\ .
\end{equation}
Here $\epsilon_{F}$ is the Fermi energy of bulk copper that is computed within extended H\"{u}ckel theory.

\begin{figure}[ht]
\begin{center}
\includegraphics[scale = 0.40]{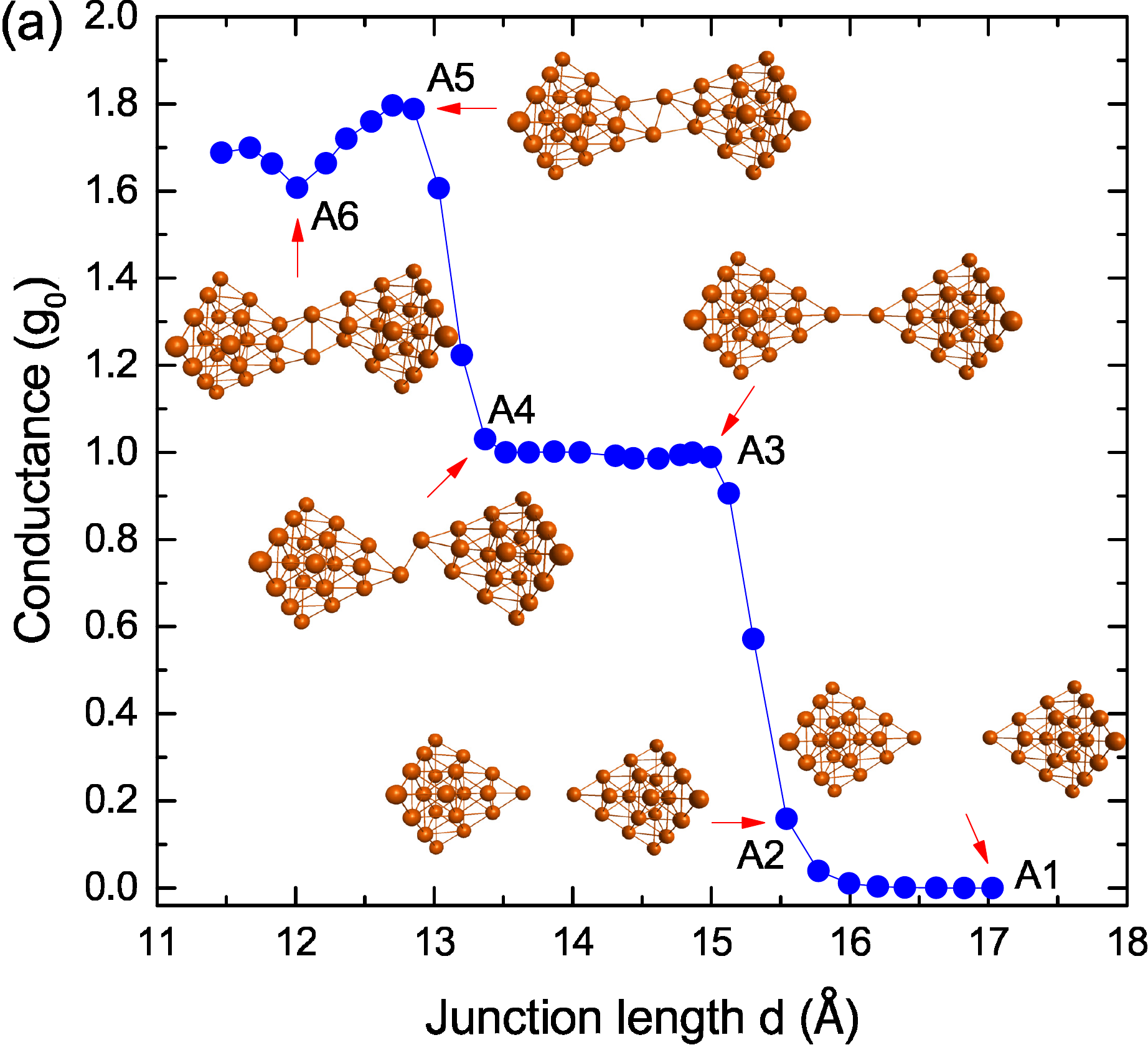}
\\
\vspace{0.05in}
\includegraphics[scale = 0.40]{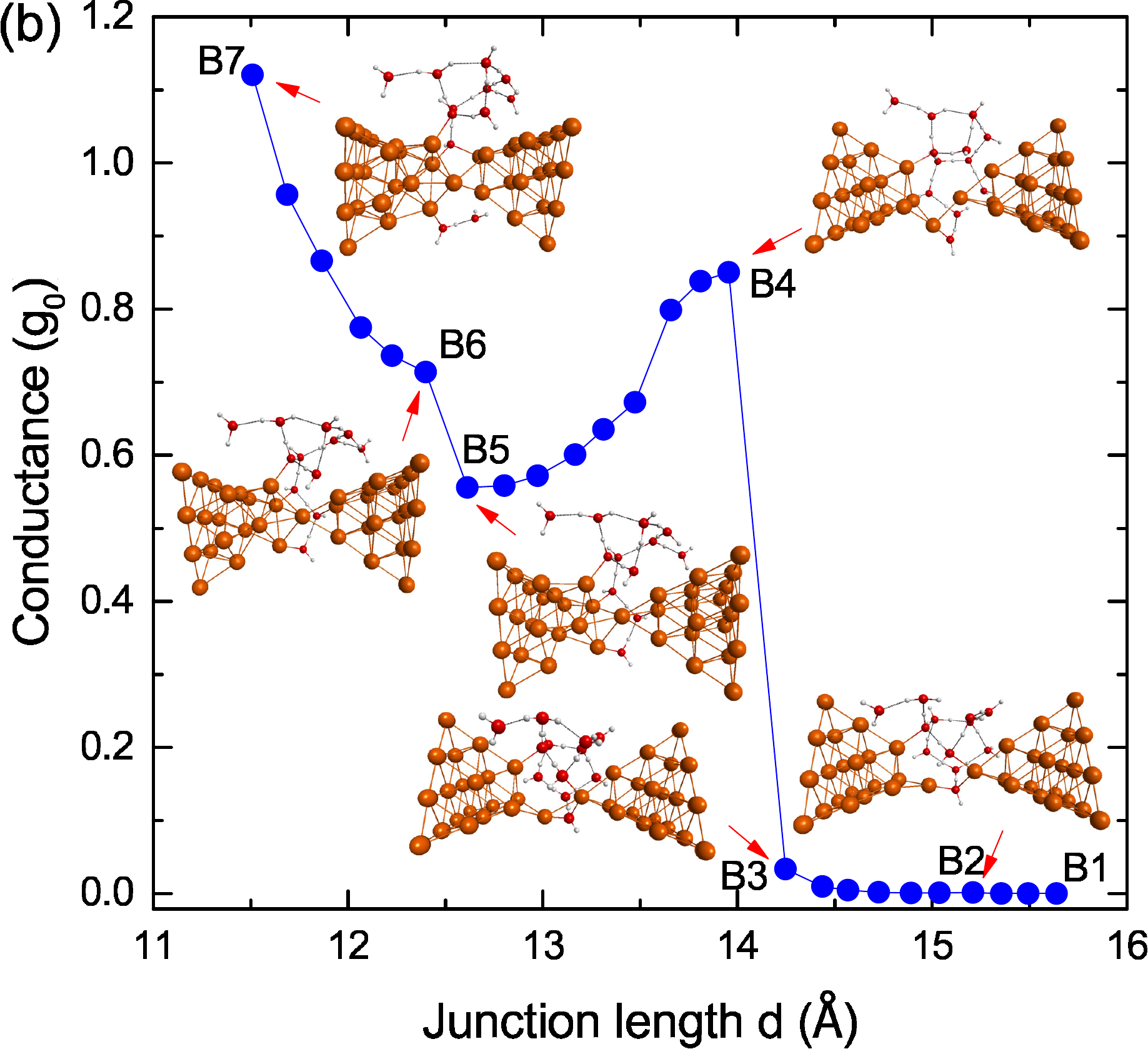}
\caption{(Color online) Calculated conductance values of a copper atomic junction with pyramidal electrodes 
in the (a) absence and (b) presence of a group of H$_{2}$O molecules as a function of junction length $d$ that is
defined as the separation between the atomic layers of the two electrodes that are furthest from the junction. 
The insets show the major structural evolutions in the atomic geometries of the junctions. }
\label{F2}
\end{center}
\end{figure}

\section{Results}

The electronic transport through the atomic junctions is strongly dependent on the electronic states of the electrode tip atoms. Also, the atomic geometry of the electrodes and their structural symmetry may affect the density of states of the electrodes. 
To show the influence of different atomic configurations of the electrodes on the tip atom electronic structure, we have depicted in Fig. 1(e) and (f) the energy dependence of local density of states of single electrodes with pyramidal and non-crystalline structures at tip atom in the absence and presence of water molecules, respectively. Among the 4$s$, 4$p$ and 3$d$ orbitals of copper atoms, the 4$s$ orbitals have the most important contribution to the tip states affecting the electron transport, while the density of states of 4$p$ and especially 3$d$ orbitals at Fermi energy is small in the given energy window. The changes of density of states with electron energy are almost the same for both structures in the absence of water molecules, which may indicate that in this case the geometry of the electrodes has a moderate effect on the density of states (see Fig. 1(e)). The similarity in the behavior of the electronic states versus energy comes from the structural details around the tip atoms. Although the geometry of the electrodes is different, the tip atom in both cases is located in a hollow site, resulting in the similar features seen in the local density of states spectra. The presence of water molecules, however, induces a significant difference between the densities of states of the 4$s$ orbitals of the two structures around Fermi energy, as shown in Fig. 1(f). In the case of pyramidal electrodes, the water molecules form a single cluster that bonds to the tip atom, while in the case of non-crystalline structure, the molecules do not show such a tendency to group together and no water molecule forms a bond with the tip atom. 
Note that for such electrodes in the atomic junctions, the hollow sites are modified as the tip atoms are brought into contact. This will be discussed below with and without introducing the water molecules in our optimized structures.

We start by considering copper atomic junctions with pyramidal electrodes before introducing water molecules. In the first optimized structure (see the inset of Fig. 2(a) at the position marked A1), the junction length, defined as the distance between the outermost atomic layers, is $d=17.03$\AA\  and the separation of the tip atoms is 5.02\AA. In this case there is negligible tunneling of electrons from one electrode to the other. At $d=15.54$\AA\ (marked A2), the tip separation is 3.36\AA\ and electrons can tunnel through the potential barrier between the two electrodes. The tunneling probability increases smoothly as the electrodes come closer together and eventually, when the junction length reaches $d=14.99$\AA\ (marked A3) with tip separation of 2.74\AA, a single-atom contact forms between the Cu tip atoms and the conductance becomes $\sim 1g_{0}$. In this case decreasing the length  of the junction further affects the conductance value very little until $d\sim 13.4$\AA\ is reached, and hence a plateau of length $\sim$ 1.6\AA\ forms.  
Since the reduction of the junction length from A3 to A4 does not form a new bond between the two electrodes and causes only a rotation in the bond between the tip atoms, the number of conducting channels through the atomic contact does not change. Note that due to the spherical nature of 4$s$ orbitals and their dominant contribution to the tip density of states (see Fig. 1(c)), the bond rotation does not affect the conductance value significantly. 

Further reduction in the junction length from A4 to A5 results in another smooth transition to another (small) plateau indicating the formation of new  inter-atomic bonds as can be seen in the inset of Fig. 2(a) at A5. As $d$ is reduced further, the  tips of the two pyramidal electrodes continue to merge. This involves bond breaking and new bond formation, as is seen in the inset of Fig. 2(a) labelled A6. This localized deformation of the atomic bonding and structural distortion by reduction of junction size from A5 to A6 are the main reasons for there being no clear plateau in the conductance profile for  $ d < 12.85$\AA.
\begin{figure}[ht]
\begin{center}
\includegraphics[scale = 0.40]{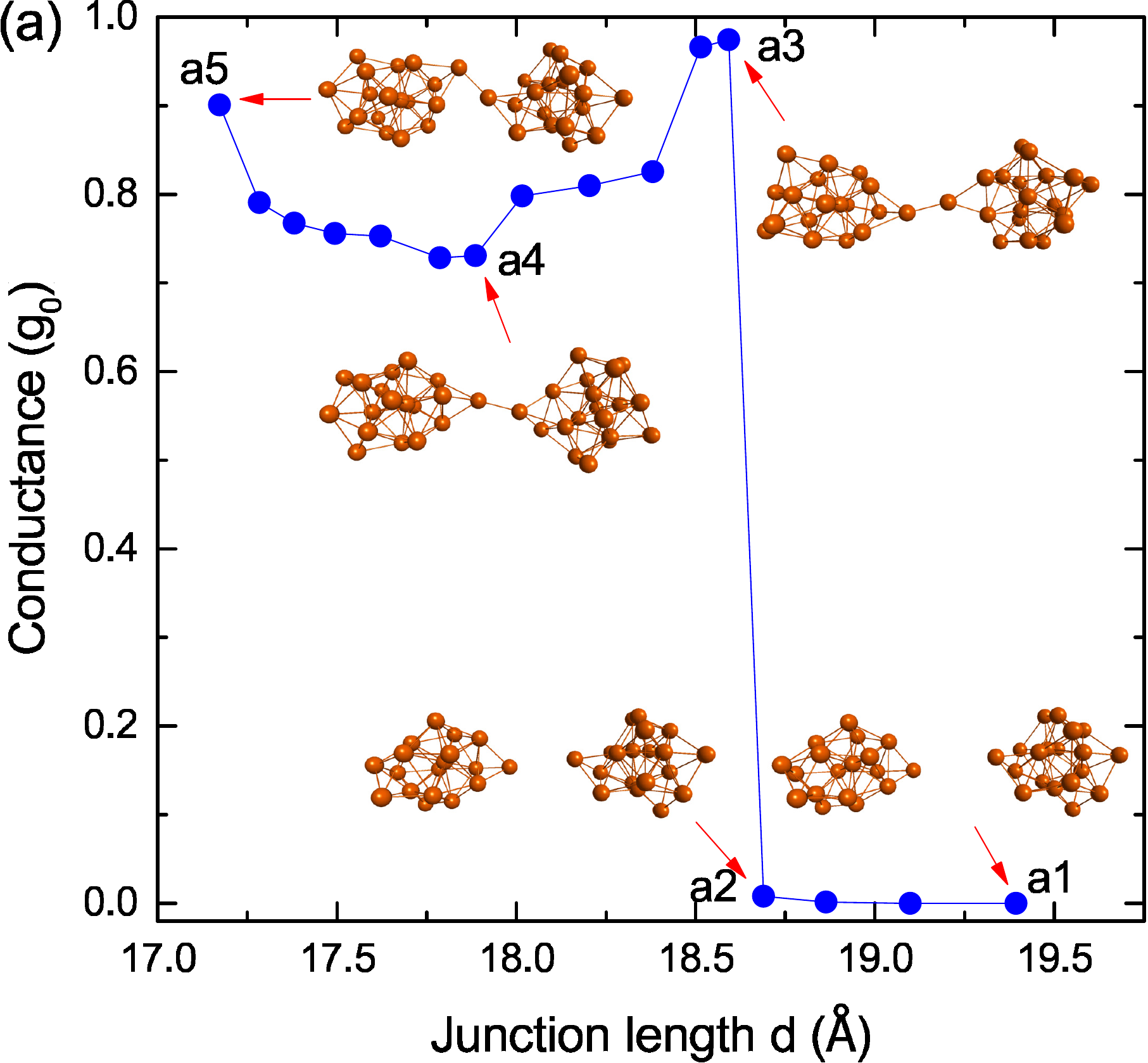}
\\
\vspace{0.05in}
\includegraphics[scale = 0.40]{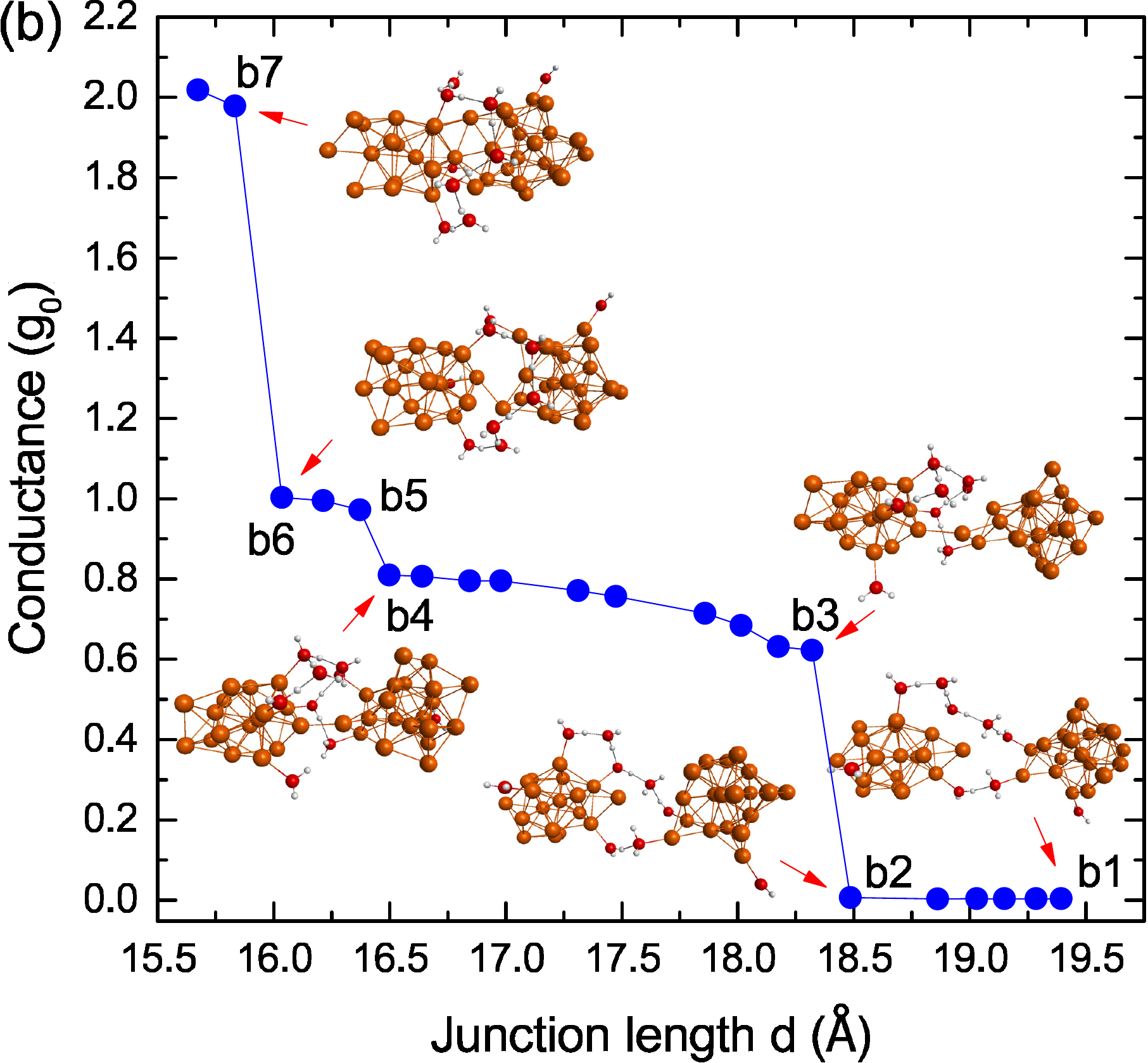}
\caption{(Color online) Calculated conductance values of a copper atomic junction with non-crystalline electrodes 
in the (a) absence and (b) presence of a group of H$_{2}$O molecules as a function of junction length $d$. 
The insets show the major structural evolutions in the atomic geometry of the junctions. }\label{F3}
\end{center}
\end{figure}

 How the electronic transport through the copper junctions with pyramid-shaped electrodes is affected by the presence of water molecules, may be seen by comparing Figs. 2(a) and 2(b). We introduced water molecules in the space between the tip atoms when the electrodes were far from each other at $d=15.64$\AA, marked B1 in Fig. 2(b). As the electrodes were gradually brought towards each other, the water molecules clustered in the gap between the two electrodes and made contact with the copper atoms in the tip regions. The tendency of hydrogen-bonded cluster of water molecules that formed between the tips was to repel the electrodes as they were brought towards each other. This resulted in deformation of the Cu atomic tips as they were brought together, and eventually the breaking of a Cu-Cu bond near the tip of the left electrode as can be seen in the inset of Fig. 2(b) at $d=15.21$\AA, marked B2. In inset B2, a single water molecule bridges the two tip Cu atoms preventing a Cu-Cu bond from forming between them. A water molecule bridging the gap between two electrodes has also been proposed in relation to a recent MCBJ experiment with Pd
electrodes \cite{Fukuzumi2018}. Although this water molecule-mediated contact is maintained up to $d=14.25$\AA  ~(inset of B3 in Fig.2(b)), the conductance remains low there because of the weakly conducting H$_{2}$O molecule bridging the two tip Cu atoms. Then, a jump to contact occurs at $d=13.95$\AA\ when one of the Cu-H$_{2}$O bonds bridging the tip atoms breaks, as shown at B4. The resulting geometry, however, is not stable when the junction length is reduced further. Then a tip atom belonging to the left electrode is pushed out of the junction onto that electrode's surface resulting in the conductance decreasing from $0.85 g_{0}$ at B4 to $0.56 g_{0}$ at B5. When $d$ decreases still further from B5 to B6, the hollow site of the right electrode deforms into a bridge site and an additional bond forms between the tip atom and the next atomic layer of the left electrode causing a jump in conductance value from 0.56$g_{0}$ to 0.71$g_{0}$. With further reduction in the junction length the conductance value  increases continuously as a result of new bond formation in the left electrode, as can be seen in the inset at B7. It is clear that in all insets of Fig. 2(b), the water molecules tend to stick to each other by forming hydrogen bonds, preventing the H$_{2}$O molecules from intercalating between the copper atomic layers of the electrodes.

That the water molecules hinder the closing of the gap between the copper electrodes can be confirmed if one compares the junction length $d=14.99$\AA\ (marked A3) at which the conductance value reaches 1$g_{0}$ in Fig. 2(a) with that ($d=13.95$\AA\ marked B4) at which a jump to contact occurs in the presence of water molecules in Fig. 2(b). The $\sim$ 1\AA\ shorter length of the copper atomic junction in the presence of water molecules reveals the repulsive effect of H$_{2}$O molecules on the copper tip atoms when the  two copper electrodes are brought together. Moreover, the absence of a pronounced conductance plateau in Fig. 2(b) suggests that the water molecules make the geometry of the tips and the arrangement of their neighboring atoms unstable causing some Cu-Cu bond breaking. 

We shall now focus on non-crystalline structures in which only the two outermost atoms are frozen. The tunneling probability of electrons between the two electrodes in the absence of water molecules remains very small while the junction length is reduced from a1 to a2 in Fig. 3(a). As the tip atoms are brought slightly closer together, a sudden increase in the conductance to $\sim 0.97g_{0}$ takes place as a result of the formation of a single-atom contact at $d=18.59$\AA, marked a3. This jump in the conductance occurs because of a bond breaking that rearranges the right tip atom from a hollow site to a bridge site at a3, thus allowing the right tip atom to abruptly move closer to the left tip. It can be seen that a small plateau has formed around $\sim 0.97g_{0}$. Since the structure of the right tip is not stable when the tips are brought closer together, the plateau is short. Upon decreasing the junction length further, the conductance drops, forming two additional short plateaus around 0.81$g_{0}$ and 0.73$g_{0}$. As shown in the inset of Fig. 3(a) at a4, the bridge site occupied by the right tip atom has deformed again to a hollow site. The hollow site geometry of both tip atoms persists as $d$  decreases further from a4 to a5 where the conductance has a rising trend. For still shorter junction lengths (not shown here), the two electrodes merge and the tip areas deform strongly.    

In contrast to the junctions with pyramidal electrodes that show a smooth transition from tunneling to a metallic contact in their conductance profile followed by a long plateau (see Fig. 2(a)), the junctions with non-crystalline structures exhibit a sudden increase in the conductance (i.e., a jump to contact) with a short plateau (see Fig. 3(a)). This reveals that the atomic arrangement of the electrodes has a crucial impact on the conductance profile of the copper electrodes in the absence of water molecules, suggesting experimental investigations in this context.

The calculated conductance for each relaxed structure with non-crystalline electrodes in the presence of water molecules is plotted in Fig. 3(b). Due to the presence of multiple water molecules in the gap between the two copper clusters, the electrodes are only connected to each other through groups of hydrogen bond-linked H$_{2}$O molecules when the Cu clusters approach each other from $d=19.39$\AA\ to 18.49\AA, labelled b1 and b2, respectively, as shown in the inset of Fig. 3(b). In this case, the probability of electrons passing through the junction is negligible. A slight further decrease in $d$ results in a jump from tunneling to a single Cu-atom contact regime with the conductance value $\sim 0.62g_{0}$ and the atomic configuration shown in the inset of Fig. 3(b) at b3. The bonding deformation of right tip atom from the hollow site to a bridge site and also the absence of a water molecule bridging two Cu tip atoms are the main factors in the sudden conductance increase from b2 to b3. It can be seen that a quasi-plateau that smoothly increases from b3 to b4 has formed, enhancing the conductance value gradually to $\sim 0.81g_{0}$ as a result of shortening the distance between the tip atoms from 2.64\AA\ to 2.49\AA. As the electrodes are brought still closer together from b4 to b7 a second jump in the conductance at b5 with a short plateau and a third jump from b6 to b7 with the conductance value increasing from $1g_{0}$ to $\sim 1.97g_{0}$ occur. In this case the contact between the two clusters evolves from a single-atomic contact at b6 to a multi-atomic contacts at b7.     
  
Comparing the conductance profiles in Fig. 3(b) and Fig. 2(b), we find some remarkable differences, revealing the role of structural geometry of copper junctions in the presence of H$_{2}$O molecules in formation of conductance plateaus. Water molecules impose several plateaus-like features in the conductance profile of junctions with non-crystalline electrodes, whereas the plateaus almost disappear in the junctions with pyramidal electrodes. The onset of the transition from tunneling to the contact regime is dependent on the geometry of the electrodes. In the junctions with the pyramidal structure, the transition takes place at $d=13.95$\AA, while it occurs at $d=18.32$\AA\ when the non-crystalline electrodes are considered.

\begin{figure}
\centerline{\includegraphics[width=1.05\linewidth]{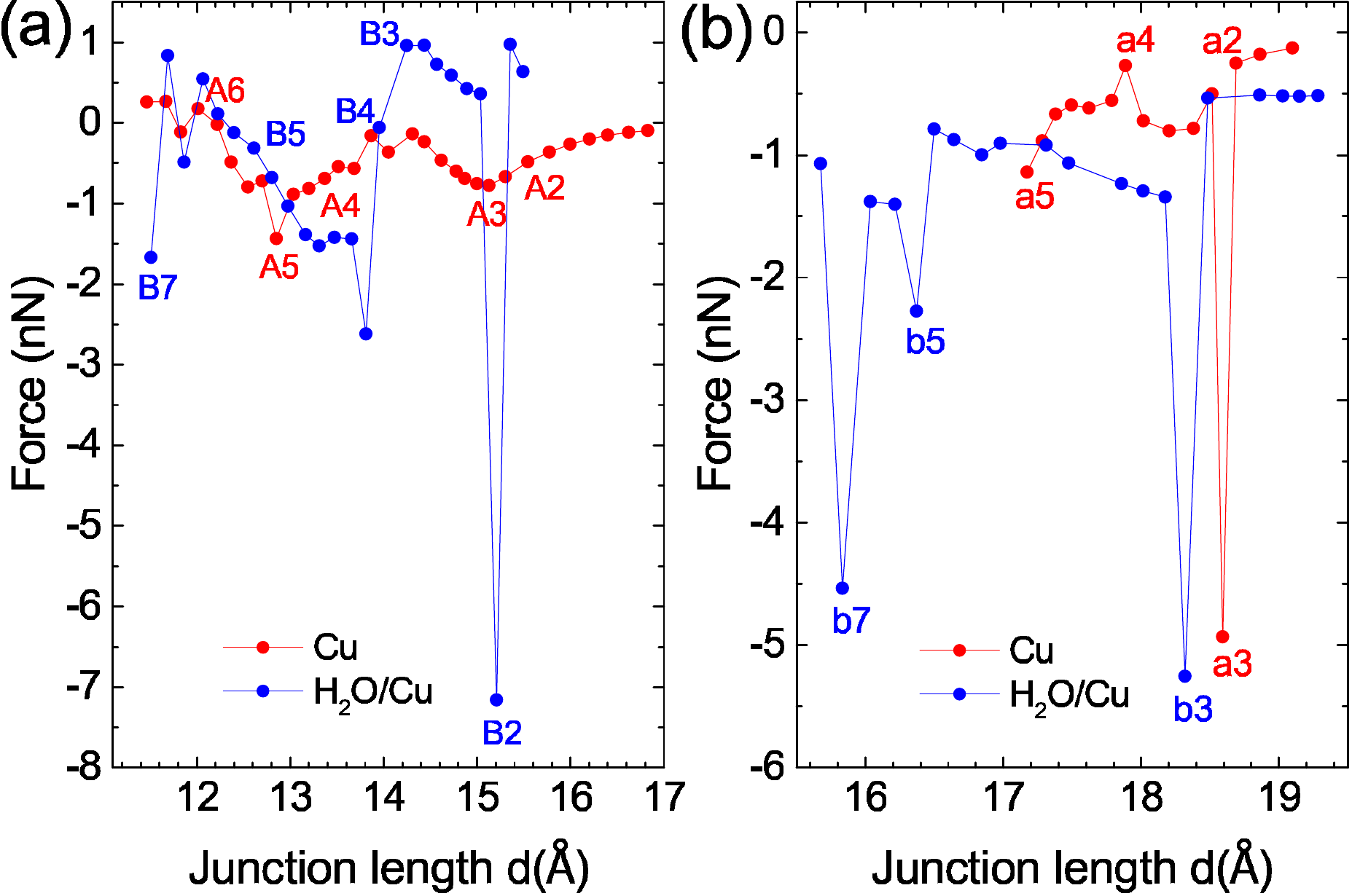}}
\caption{(Color online) Calculated the mechanical force required to bring the copper atomic junctions into contact in the case of (a) pyramidal and (b) noncrystalline electrodes before and after introduction of water molecules. The labels are the same as those in Figs. \ref{F2} and \ref{F3}.} \label{F4}
\end{figure}

It should be emphasized that when the non-crystalline electrodes with water molecules approach each other, no single H$_{2}$O molecule bridges the copper electrodes. Instead, a group of water molecules through hydrogen bonding bridge the electrodes, as shown in the inset of Fig. 3(b) at b1 and b2. By contrast, for junctions with pyramidal electrodes a single H$_{2}$O molecule bridges the copper electrodes, as shown in the inset of Fig. 2(b) at B2. Behavior similar to the latter has also been proposed in H$_{2}$O/Pd junctions \cite{Fukuzumi2018}, H$_{2}$/Pt junctions \cite{Djukic2006,Christlieb2002,Djukic2005,Thygesen2005} and in Au junctions in the presence of alkanedithiolate molecules \cite{George2010} in which a molecule is often suspended between two electrodes, while the junction length is varied.

The mechanical force, $F_{m}$, in the process of jump to contact can be calculated by means of numerical derivative of the total energy with respect to the junction length $d$ using the equation $F_m(d_{i})=-[E(d_{i+1})-E(d_{i})]/\delta_i$ where $d_{i}$ is the length parameter of compression step $i$, $E(d_{i})$ is the total energy in terms of $d_{i}$, and $\delta_i=d_{i+1}-d_{i}$. We have depicted in Fig. 4 the magnitude of the force vs parameter $d$ for both types of electrodes before and after the introduction of water molecules. 

The force is mostly negative (attractive) with magnitudes less than 2nN in the case of pyramidal electrodes without water molecules, while the force can be positive (repulsive) or negative (attractive) in the presence of water molecules along with a single sharp spike, labelled B2, at the onset of bridging the two electrodes by a single water molecule (see Fig. 4(a)). This means that the bond formation between two pyramidal electrodes in an aqueous environment manifests itself as a short ranged strong attraction between the two electrodes during the jump to contact. In the case of noncrystalline electrodes (see Fig. 4(b)), however, the force is always negative whether in vacuum or in the presence of water molecules. The strong spikes in the force values appear when a jump to contact or the formation of a new plateau occurs. Comparing the conductance profiles of Fig.3 with Fig.4(b), it is seen that the force shows a single strong spike at a3 when a jump to contact happens for electrodes in vacuum, whereas there are three spikes at b3, b5, and b7 when water molecules are present. The strong spike at b3 corresponds to the jump to contact, while the labels b5 and b7 correspond to the formation of plateaus in the conductance profile. For both pyramidal and noncrystalline electrodes the jump to contact in the presence of water molecules results in a conductance of approximately $0.8g_{0}$ in marked contrast to the behavior in the absence of water molecules where the first plateau forms at approximately $1g_{0}$ in both the pyramidal and noncrystalline cases.

Therefore, as can be seen in Figs. 3 and 4, the conductance and force can each signal the presence of water molecules in copper atomic junctions experimentally. Hence, in addition to being of fundamental interest, these findings may be relevant for nanoelectronic applications.
 
\section{Discussion}

In summary, based on a combination of \textit{ab initio} and semi-empirical calculations, we have presented a systematic
exploration of coherent transport through copper atomic junctions with pyramidal and non-crystalline electrodes in the presence and absence of water molecules. The junction-size dependence of conductance as the two copper electrodes approach each other before introduction of H$_{2}$O molecules shows a smooth transition from tunneling to a single-atom contact regime with a conductance plateau at 1$g_{0}$ for junctions with pyramidal electrodes, while it exhibits a jump to contact with a short conductance plateau when non-crystalline electrodes are utilized. In the presence of a group of water molecules, both junctions with pyramidal and non-crystalline electrodes display a sudden jump to contact transition from tunneling to metallic conduction in their conductance profiles. These profiles exhibit several plateaus for the junctions with non-crystalline electrodes but only a short quasi-plateau in the pyramidal junctions. We predict that bonding between a single H$_{2}$O molecule and the two metallic tip atoms with the molecule bridging the two Cu electrodes can only form in the junctions with pyramidal electrodes as a result of structural symmetry of the junctions and additional constraints on the Cu atoms. 
 
We show that the mechanism of the jump to contact (or its absence) is structure dependent which can be understood as follows: The pyramidal electrodes exhibit no jump to contact in vacuum. They come into contact as a result of gradual bond formation with the bond length between the tip atoms decreasing continuously as the junction is narrowed. This manifests itself as a smooth conductance transition from tunneling to the contact regime followed by a wide $g=1g_{0}$ plateau. The electrodes retain their pyramidal symmetry throughout this process. The mechanism is different in the case of noncrystalline electrodes in vacuum.  These electrodes do not have a specific symmetry and the copper atoms are relatively free to rearrange. Thus, the tip atoms come abruptly into contact through a process of bond breaking that accompanies the formation of a new bond between the electrodes. This manifests itself as a strong spike in the profile of mechanical force (see a3 in Fig. 4(b)) and an abrupt transition from a low conductance to $g\simeq1g_{0}$ in Fig.3(a). For the case of electrodes in an aqueous environment, initially a single water molecule bridges the two pyramidal electrodes. This is detectable as a strong spike in the force profile (see B2 in Fig. 4(a)), while the conductance remains low. Then, the bond between one of the copper tip atoms and the bridging H$_{2}$O molecule breaks and a direct bond forms between the two tip copper atoms. This is a jump to contact event in which the conductance increases abruptly from a low value to $g\simeq0.8g_{0}$ followed immediately by the force between the electrodes becoming strongly attractive.  The water molecules mostly stay together in the vicinity of the tip region, due to hydrogen bonding. By contrast, for the noncrystalline electrodes, the water molecules make hydrogen bonded quasi-molecular chains connecting the two electrodes instead of a single molecule bridging the two tip atoms. Then, one molecular chain breaks, opening the way for a direct copper-copper bond to form between the two electrodes. This direct bond formation manifests itself as a jump to contact event in which the conductance increases abruptly from a low value to $g\simeq0.62g_{0}$ and an attractive spike is seen (b3 in Fig. 4(b)) in the force between the electrodes.

Our results suggest that the tip atoms of the pyramidal junctions are much more stable than those in the junctions with non-crystalline electrodes before H$_{2}$O  molecules are introduced, while the tip atoms in the presence of water molecules are more stable in the noncrystalline junctions than those in the junctions with pyramidal electrodes. These findings reveal the importance of geometrical structure of tip electrodes in the atomic junctions and also the interplay between these electrodes and water molecules which may exist in the conductance measurement apparatus. 

In all of the cases studied in this paper, both electrodes initially terminate in single Cu atoms, each of which is bound to three nearest neighbor Cu atoms. The present work has revealed that, even for such superficially similar structures of the tips of Cu electrodes, whether the jump to contact phenomenon occurs or not (and its physical mechanism) depend qualitatively on the details of the atomic structure of the metal electrodes further from the tip atoms. These structural details affect the jump to contact phenomenon strongly when the tips of the electrodes are brought together in vacuum. They also affect strongly how the presence of water molecules in the junction influences the jump to contact phenomenon. It appears reasonable to expect the mechanisms of the jump to contact phenomenon to be similarly subtle for other atomic terminations of the metal electrodes. Therefore, further theoretical and experimental studies of this remarkably rich and subtle phenomenon are clearly desirable.   

\section*{Acknowledgement}
This work was supported by NSERC, WestGrid, and Compute Canada.

\end{document}